\documentclass[journal]{IEEEtran}
\usepackage{graphicx}
\usepackage{epsf}

\usepackage{amsmath}
\usepackage{amssymb}
\usepackage{amsthm}
\usepackage{latexsym}
\usepackage{multirow}
\usepackage{multicol}
\usepackage[table]{xcolor}
\usepackage[hidelinks]{hyperref}
\usepackage{algorithm2e}
\usepackage{subcaption}

\usepackage{mathtools}

\begin{document}

\title{{\fontsize{20}{4}\selectfont Prioritization of Basic Safety Message in DSRC Based on Distance to Danger}}

\author
{
Seungmo Kim, \textit{Member}, \textit{IEEE}, and Byung-Jun Kim

%\vspace{-0.4 in}

\thanks{S. Kim is with the Department of Electrical and Computer Engineering, Georgia Southern University in Statesboro, GA, USA. B. J. Kim is with the Department of Statistics, Virginia Tech in Blacksburg, VA, USA.}
}

\maketitle

% Abstract
%%%%%%%%%%%%%%%%%%%%%%%%%%%%%%%%%%%%%%%%%%%%%%%%%%%%%%%%%%%%%%%%%%%%%%%%%%%%%%%%%%%%%%%%%%%%%%%%%%%%%%%%
\begin{abstract}
Many parties claim the technical significance of Dedicated Short-Range Communications (DSRC) in intelligent transportation system (ITS) for promotion of transportation safety. The main challenge in this key vehicle-to-everything (V2X) standard is high odds of network congestion. Furthermore, in accordance with a V2X network being inherently dynamic in key aspects such as vehicle density and velocity, the networking behavior of a DSRC system is usually highly complicated to analyze. In addition, the United States Federal Communications Commission (US FCC) recently proposed the so-called ``5.9 GHz band innovation,'' which reduced the dedicated bandwidth for DSRC to 10 MHz at best from 75 MHz. Motivated from these challenges, the necessity of ``lightening'' the networking load of a DSRC network has become essential to keep safety-related operations from performance deterioration. To this end, this paper proposes a protocol that prioritizes transmission of a basic safety message (BSM) at a vehicle according to the level of accident risk of the vehicle. The proposed protocol uses the distance of a vehicle from a danger source as the metric to determine the priority for transmission. Our results show that this protocol effectively prioritizes the transmission opportunity to dangerous vehicles, and hence results in higher performance in terms of key metrics--i.e., average delay, throughput, and inter-reception time ($\mathsf{IRT}$).
\end{abstract}


\IEEEpeerreviewmaketitle

%%%%%%%%%%%%%%%%%%%%%%%%%%%%%%%%%%%%%%%%%%%%%%%%%
\section{Introduction}
\subsection{Background}
\subsubsection{5.9 GHz Band for V2X}
Vehicle-to-Everything (V2X) communication has the potential to significantly bring down the number of vehicle crashes, thereby reducing the number of associated fatalities \cite{gdot_1}. The capability gave V2X communications the central role in constitution of intelligent transportation system (ITS) for connected vehicle environments. Today, two key radio access technologies (RATs) that enable V2X communications have been attracting the greatest research interest: namely, DSRC and cellular V2X (C-V2X). DSRC was designed to primarily operate in the 5.9 GHz band (viz., 5.850-5.925 GHz) ever since the band was dedicated in the United States by the Federal Communications Commission (FCC) in 1999. However, C-V2X has recently been proposed to operate in the 5.9 GHz band in addition to the cellular operators' licensed carrier \cite{gdot_2}.

\subsubsection{Significance of DSRC}
Of the two RATs, the Dedicated Short-Range Communications (DSRC) has longer been deployed in many communities for safety-critical applications owing to several merits. First, the most important benefit for advocating DSRC as the key enabler of V2X communications is that it is a proven technology: it has been tested by car manufacturers for more than 10 years. {\color{black} Second, DSRC does not require any paid subscription, which makes possible wider deployment at a lower cost. Third, the universal compatibility among IEEE 802.11 technologies leads to the spectrum versatility and easy operation, which could strengthen DSRC in the market of connected vehicles.}

Based on these advantages, as of November 2018, more than 5,315 roadside units (RSUs) operating in DSRC were deployed nationwide \cite{gdot_3}. In December 2016, the National Highway Traffic Safety Administration (NHTSA) proposed to mandate DSRC for all new light vehicles \cite{gdot_4}. However, despite the advantages and widespread deployment, the technology has just encountered the biggest obstacle ever: the \textit{5.9 GHz band reallocation} by the US FCC \cite{nprm}.

\subsubsection{5.9 GHz Band Reform by the US FCC}
Out of the 75 MHz of bandwidth in the 5.9 GHz band (i.e., 5.870-5.925 GHz), in December 2019, the US FCC voted to allocate the lower 45 MHz (i.e., 5.850-5.895 GHz) for unlicensed operations to support high-throughput broadband applications (e.g., Wireless Fidelity, or Wi-Fi) \cite{nprm}. While the reallocation is proposing to leave the upper 30 MHz (i.e., 5.895-5.925 GHz) for ITS operations (i.e., DSRC and C-V2X), it is also proposing to dedicate the upper 20 MHz of the chunk (i.e., 5.905-5.925 GHz) for C-V2X.

Therefore, according to this plan, at best, DSRC is only allowed to use 10 MHz of spectrum (i.e., 5.895-5.905 GHz). It has \textit{never been studied nor tested if a 10 MHz would suffice} for operation of the existing DSRC-based transportation safety infrastructure. Many states in the US have already invested large amounts of fortune in the deployment of connected vehicle infrastructure based on DSRC \cite{gdot_6}. As such, it has become urgent to understand how much impact of the FCC's 5.9 GHz band reallocation will be placed on the performance of such connected vehicle infrastructure.

\subsubsection{Necessity of Lightening DSRC Networking Load}
Not only the significantly smaller bandwidth, DSRC may need to experience coexistence with C-V2X users according to the FCC's proposition \cite{nprm}. The key technical challenge here is that the C-V2X standards adopt significantly different protocols, which makes the technology incompatible with DSRC-based operations. Based on the author's recent investigation \cite{arxiv19}, on average 4.63 C-V2X users can corrupt a DSRC packet if the two disparate RATs operate in a 10-MHz channel. It implies that C-V2X-to-DSRC interference may occur very often, considering such a small number of (viz. less than 5) C-V2X users to corrupt a DSRC packet.

Therefore, it has become crucial to lighten the load of a DSRC network while keeping the dissemination of packets operable, in order to suit the technology into such a competitive spectrum environment.

\subsection{Contributions}
As shall be detailed in Section \ref{sec_related}, the current literature shows a key limitation to achieve the load lightening of a DSRC network: despite being a predominant factor determining the performance of a V2X network, the length of backoff time was allocated to each vehicle without considering ``semantic'' contexts.

We would regard it more efficient from the system's point of view if vehicles being closer to a danger take higher chances to transmit. The rationale is that these initial basic safety messages (BSMs) will propagate through the network, which will eventually make most of the vehicles in the network able to receive the BSMs and hence promote the level of safety.

To this line, this paper proposes a V2X networking scheme where a vehicle takes a transmission opportunity according to the probability that it runs into a crash. Moreover, we clearly distinguish our contributions from the most relevant work \cite{secon19}\cite{milcom19}. While the prior work focused on the stochastic geometry of a particular coexistence scenario between military and civilian vehicles in an urban area, this present paper significantly extends the scope of discussion to (i) a general two-dimensional geometry and (ii) detailed analysis on networking behaviors--i.e., an exact backoff allocation method.

Overall, the technical contributions of this paper distinguished from the literature can be summarized as follows:
\begin{enumerate}
\item It proposes a method prioritizing a BSM according to the level of danger to which each vehicle is exposed.
\item In order to measure the risk, it uses the ``distance to a danger source,'' which is a quantity that is easy to obtain by using the existing techniques and apparatus.
\item Based on (i) key metrics--namely, delay, throughput, and $\mathsf{IRT}$--and (ii) a generalized two-dimensional spatial model (not limited to certain road models), it provides a stochastic analysis framework characterizing a DSRC network's broadcast of BSMs.
\end{enumerate}

\begin{table}[t]
\caption{Frequently used abbreviations}
\centering
\begin{tabular}{ |l|l|}
\hline
\textbf{\cellcolor{gray!30}Abbreviation} & \cellcolor{gray!30}\textbf{Description}\\
\hline\hline
BEB & Binary exponential backoff\\
BSM & Basic safety message\\
CAT & Category of distance from danger, $d_{\rightarrow \text{dgr}}$\\
CSMA & Carrier-sense multiple access\\
C-V2X & Cellular V2X\\
CW & Contention window\\
dgr & The danger source (See Figure \ref{fig_scatter})\\
DSRC & Dedicated Short-Range Communications\\
EXP & Packet expiration\\
FCC & US Federal Communications Commission\\
HN & Collision by hidden node\\
$\mathsf{IRT}$ & Inter-reception time\\
ppd & The ``proposed'' backoff scheme\\
PPP & Poisson point process\\
STA & Station\\
SYNC & Collision by synchronized transmission\\
tdl & The ``traditional'' BEB scheme\\
V2X & Vehicle-to-everything communications\\
\hline
\end{tabular}
\label{table_abbreviations}
\end{table}

%%%%%%%%%%%%%%%%%%%%%%%%%%%%%%%%%%%%%%%%%%%%%%%%%
\section{Related Work}\label{sec_related}
\subsection{Performance Analysis Schemes}
\subsubsection{Mathematical Analysis Framework}
Analysis frameworks based on stochastic geometry for DSRC have been provided recently \cite{arxiv19}-\cite{access19}. They commonly rely on the fact that uniform distributions of nodes on $X$ and $Y$ axes of a Cartesian-coordinate two-dimensional space yield a Poisson point process (PPP) on the number of nodes in the space \cite{haenggi05}. This paper also applies the stochastic geometry framework for analysis of the proposed mechanism.

\subsubsection{Performance Evaluation Method}
A recent proposal combines a packet-level simulation model with data collected from an actual vehicular network \cite{gdot_13}. {\color{black} It is critical to discuss the potential impacts of \textit{internal} and \textit{external} bandwidth contentions, which form a critical discussion point after the US FCC's recent 5.9 GHz band reallocation \cite{nprm}. The ``internal'' contention means the contention among DSRC vehicles themselves, while the ``external'' contention refers to the contention incurred by other RAT(s).}

{\color{black} The limitation of the prior art lies in that the performance evaluation was performed without consideration of these bandwidth contentions, which might undermine its own generality.} For instance, it is assumed that {\color{black} (i) safety messages and (ii) packets for non-safety applications} are sent over separate DSRC channels \cite{gdot_13}, whereby no interference is generated between safety and Internet traffic. This assumption has become obsolete according to the US FCC's recent proposition where DSRC is unable to utilize multiple channels any more \cite{nprm}.

\subsubsection{DSRC in High Traffic Density}
{\color{black} It has been found that a DSRC network is more constrained by packet expirations (EXPs) rather than collisions over the air \cite{arxiv2005}. An \textit{EXP} refers to a packet ``drop'' as a result of not being able to (i) make it through the backoff process and hence (ii) be transmitted within a beaconing period. Since the IEEE 802.11p broadcast of BSMs does not support retry nor ACK, an expired packet is dropped and the next packet with a new sequence number is generated \cite{pimrc11}-\cite{eurasip19}. The reason of a packet not being able to go through a backoff process is finding the medium busy, which hinders the backoff counter from being decremented. Also, notice that a \textit{collision} is composed of two types of cause: a synchronized transmission (SYNC) or a hidden node (HN).}

Thus, we put particular focus on the performance of a DSRC network in a high density of traffic. The performance of a DSRC broadcast system in a high-density vehicle environment has been studied \cite{gdot_11}, yet the assumption was too ideal to be realistic--i.e., the number of vehicles within a vehicle's communication range was kept constant. Another study proposed a DSRC-based traffic light control system \cite{gdot_12}, but it limited the applicability to the traffic lights only.

\subsubsection{Safety-Related Application}
Furthermore, we concentrate on DSRC's networking to support the safety-critical applications. In the related literature, a DSRC-based end of queue collision warning system has been proposed \cite{gdot_14}. However, it discusses a one-dimensional freeway model, which needs significant improvement for application to an intersection with two or more ways.

\subsubsection{External Bandwidth Contention}
Lastly, the objective of our proposed protocol is to lighten the traffic load of a DSRC network to better suit in an environment of coexisting with a disparate technology (i.e., C-V2X) according to the 5.9 GHz reallocation \cite{nprm}. The performance degradation of DSRC under interference from Wi-Fi has been studied \cite{gdot_15}. Moreover, methods of enabling coexistence between dissimilar RATs were proposed in various other scenarios including: (i) commercial wireless and military pulsed radar at 3.5 GHz \cite{icnc}-\cite{lett}; (ii) 5G and incumbent in mmW bands \cite{jsac}-\cite{verboom}; and (iii) general cognitive radio schemes \cite{crown08}-\cite{crown11}. However, the prior work commonly lacks consideration of coexistence with C-V2X.

\subsection{Performance Improvement Schemes}
{\color{black} Various modifications on the binary exponential backoff (BEB) algorithm have been proposed as a means to improve throughput and fairness in general carrier-sense multiple access (CSMA) in IEEE 802.11-related technologies.} Specifically, adjustment of the CW was often suggested to improve the performance of a vehicular communications network such as a recent work \cite{wu2018improving}. More directly relevant to our work, a distance based routing protocol has been found to perform better in vehicular ad-hoc networks (VANETs) \cite{ramakrishna2012dbr}. Also, in a general ad-hoc network, reduction of the length of a header can be a solution that is worth considering \cite{grasnet}; however, due to being a centralized architecture, it shows a limit to be applied to a V2X network. A ``subjective'' user-end experience optimization is also worth consideration \cite{gilsoo}, wherein a one-bit user satisfaction indicator was introduced, which served as the objective function in a non-convex optimization.

\begin{table}[t]
\caption{Key notations}
\centering
\begin{tabular}{ |l|l|}
\hline
\textbf{\cellcolor{gray!30}Notation} & \cellcolor{gray!30}\textbf{Description (unit)}\\
\hline\hline
$d_{\rightarrow \text{dgr}}$ & Distance of a vehicle to the danger source (m)\\
$\lambda$ & Vehicle density (vehicles/m$^{-2}$)\\
$\mathsf{N}_{\text{bcn}}$ & Number of beaconing periods with failed packet delivery (EA)\\
$\mathsf{N}_{\text{bo}}$ & Number of slots spent during a backoff process (EA)\\
$\mathsf{N}_{\text{sta}}$ & Number of STAs competing for the medium (EA)\\
$\mathsf{R}$ & Normalized throughput\\
$\tau$ & Probability of a transmission\\
$\mathsf{T}_{\text{bo}}$ & Time length taken for a backoff process (sec)\\
$\mathsf{T}_{\text{ibi}}$ & Time length taken for a packet collision (sec)\\
$\mathsf{T}_{\text{exp}}$ & Time length taken for a packet expiration (sec)\\
$\mathsf{T}_{\text{ibi}}$ & Time length of the inter-broadcast interval (sec)\\
$\mathsf{T}_{\text{suc}}$ & Time length taken for a successful packet delivery (sec)\\
Th$_{i}$ & Threshold on $d_{\rightarrow \text{dgr}}$ for CAT $i$ (m)\\
\hline
\end{tabular}
\label{table_notations}
\end{table}

\begin{figure}
\centering
\includegraphics[width = \linewidth]{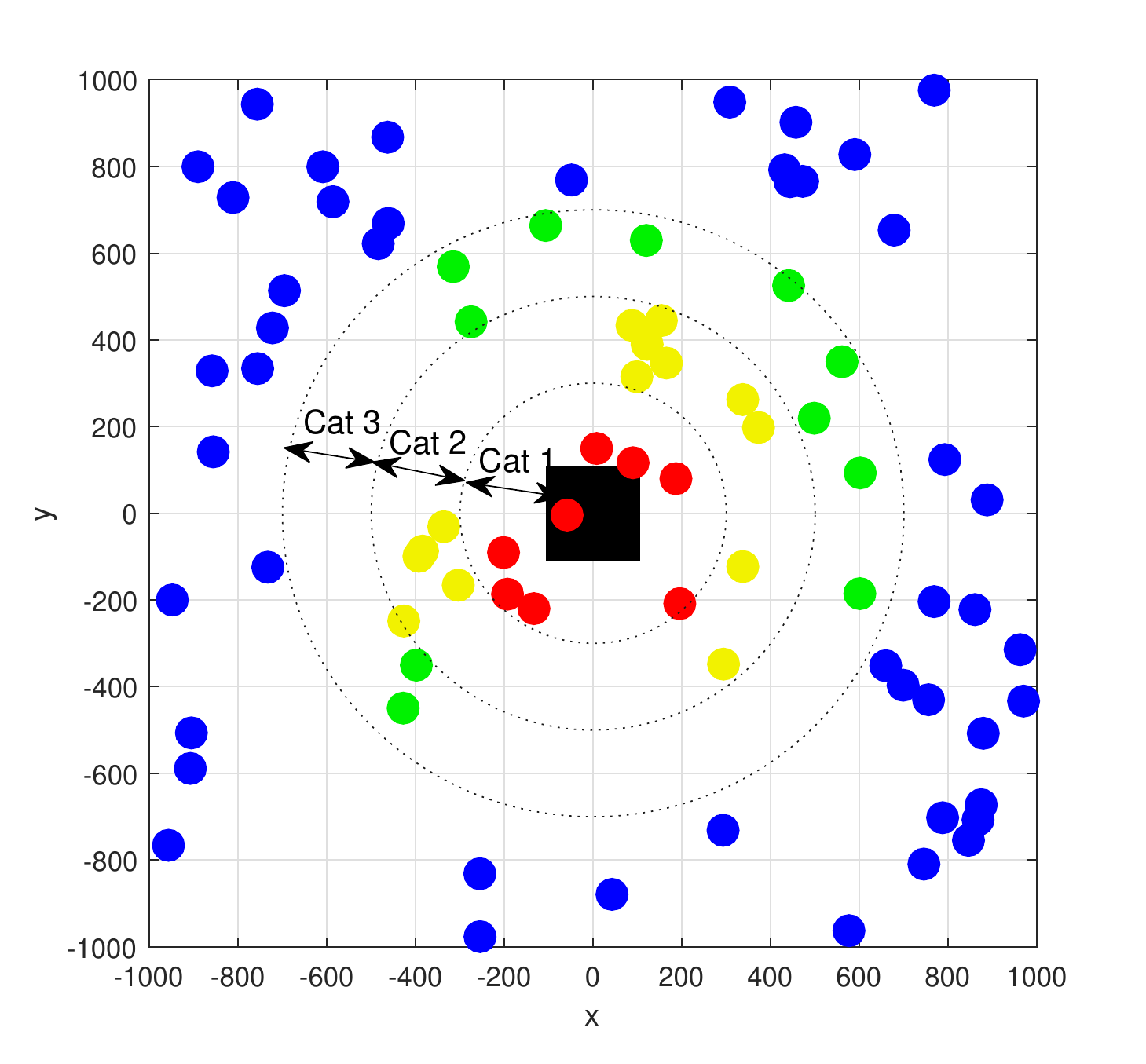}
\caption{An example drop of nodes on $\mathbb{R}^2$ (with the danger source at the origin, $\lambda = 2 \times 10^{-5}$ m$^{-2}$, CAT 1: $0 \le d_{\rightarrow \text{dgr}} \le \text{Th}_{1}$, CAT 2: $\text{Th}_{1} < d_{\rightarrow \text{dgr}} \le \text{Th}_{2}$, CAT 3: $\text{Th}_{2} < d_{\rightarrow \text{dgr}} \le \text{Th}_{3}$ where \{Th$_{1}$, Th$_{2}$, Th$_{3}$\} = \{300, 500, 700\} m)}
\label{fig_scatter}
\end{figure}

%%%%%%%%%%%%%%%%%%%%%%%%%%%%%%%%%%%%%%%%%%%%%%%%%%
\section{System Model}
This section describes the system model that this paper adopts for analysis. Note that Table \ref{table_abbreviations} lists key abbreviations that are frequently used throughout this paper. Also, mathematical notations are summarized in Table \ref{table_notations}.

\subsection{Geometry}
A two-dimensional space $\mathbb{R}^2$ is defined as a 2 km-by-2 km square, as illustrated in Figure \ref{fig_scatter}. Once a vehicle reaches the end of the space, it bounces back into the space. This assumption is to maintain a fixed vehicle density and, hence, a same level of competition for the medium at any given time.

The distribution of the nodes follows a \textit{homogeneous PPP} in $\mathbb{R}^2$. We define a general situation where a safety-critical application disseminates BSMs over a V2X network \cite{j2735}. That is, a source of ``danger'' exists (expressed as a large black square in Figure \ref{fig_scatter}), which should be avoided by all the other vehicles. The danger source is located at the origin, i.e., the very center of $\mathbb{R}^2$.

As shall be detailed in Section \ref{sec_proposed}, our proposed algorithm prioritizes transmission of a BSM as a vehicle is closer to this danger source. This necessitates to measure the \textit{distance from the danger}, which is denoted by $d_{\rightarrow \text{dgr}}$. Figure \ref{fig_scatter} demonstrates an example ``drop'' of vehicles with the density of $\lambda = 2 \times 10^{-5}$ m$^{-2}$, which is equivalent to 80 nodes over the defined space $\mathbb{R}^2$. The crash risk is categorized by using $d_{\rightarrow \text{dgr}}$ as follows:
\begin{align}\label{eq_cats}
\text{CAT 1 (``Most dangerous'')}&: 0 \le d_{\rightarrow \text{dgr}} \le \text{Th}_{1}\nonumber\\
\text{CAT 2 (``Less dangerous'')}&: \text{Th}_{1} < d_{\rightarrow \text{dgr}} \le \text{Th}_{2}\nonumber\\
\text{CAT 3 (``Far less dangerous'')}&: \text{Th}_{2} < d_{\rightarrow \text{dgr}} \le \text{Th}_{3}
\end{align}
In Figure \ref{fig_scatter}, the vehicles positioned within CATs 1, 2, and 3 are marked as red, yellow, and green circles, respectively.

The vehicles that are sufficiently far are drawn as blue circles. As shall be depicted in Section \ref{sec_proposed}, the proposed protocol does not allocate these vehicles not belonging to any of the three CATs. The rationale behind this is that these farthest located vehicles are within communications ranges of those belonging to CAT 3. That is, once vehicles in CAT 3 become able to transmit, the messages can be disseminated to these even further vehicles.

{\color{black} Notice, though, that the categorization of CW into the three chunks is an example as an effort to show how the entire protocol works. In other words, of course there could be more than the three CW ranges. This paper provides a general framework, which can always be extended to dividing into \textit{a larger number of ranges}. As such, the division into three ranges used in this paper does not represent the focal point of this paper.}

\subsection{Communications}
We suppose that all the vehicles distributed in $\mathbb{R}^2$ have the same ranges of carrier sensing and communication. Also, each vehicle broadcasts a BSM every 100 msec, which is denoted by $\mathsf{T}_{\text{ibi}}$--\textit{i.e.}, 10 Hz of the broadcast rate.

We remind that DSRC adopts distributed coordination function (DCF) as the basic access mechanism \cite{11pstd}. This paper assumes that the DCF operates in a saturated-throughput scenario \cite{bianchi}. The purpose of this assumption is to analyze a worst-case scenario (i.e., the heaviest possible network load), which can provide a conservative guideline for the performance evaluation of the proposed scheme in a DSRC network.

Lastly, in accordance with the FCC's 5.9 GHz reallocation \cite{nprm}, this paper assumes operation of DSRC in only one channel being 10 MHz wide.

%%%%%%%%%%%%%%%%%%%%%%%%%%%%%%%%%%%%%%%%%%%%%%%%%%
\section{Proposed Algorithm}\label{sec_proposed}
We propose a protocol that controls the value of $k$ to prioritize transmission by a vehicle at a higher crash risk as a means to improve the best knowledge in the literature and suit to supporting safety-critical messaging in DSRC, as has been discussed in Section \ref{sec_related}.

\subsubsection{Key Improvement from Conventional CSMA}
It has been noted that for contention resolution, the conventional BEB algorithm relies on the number of unsuccessful transmission attempts and Physical Layer (PHY) related constant values including packet retry limit, maximum and minimum values of CW size, header format, etc \cite{patel15}. This specifically means that once the PHY specific values are fixed, the future course of the BEB algorithm would be dictated by the number of unsuccessful attempts taken by a STA to successfully transmit the packet. {\color{black} In fact, the PHY parameters will likely remain constant since the current version of IEEE 802.11p does not support a link adaptation \cite{stdmag19}\cite{sensors19}.}

We got motivated from the curiosity of why it should be mandatory to have a uniform probability of choosing a backoff time for all Txs competing for the medium. Practically, if we have 50 vehicles on the road at a certain time instant and if they try to transmit a packet at the same time, all of them will have an equal opportunity to choose for a backoff time randomly from a range of $[0,\text{CW}]$. Regardless of how close or far the Tx vehicle to a danger area is, a vehicle will choose the backoff time in a random manner, while which it will have to hold the transmission. For instance, the Tx STA being far away from the danger source (and thus at a lower risk of a crash) can still be allocated a shorter backoff time. Therefore, here we propose an idea of assigning a backoff time depending on $d_{\rightarrow \text{dgr}}$, the distance between a STA and the danger source. Specifically, a Tx STA with a smaller $d_{\rightarrow \text{dgr}}$ (i.e., closer to the danger source) will have a shorter backoff time and vise versa.

\begin{figure}[t]
\hspace{-0.2 in}
\includegraphics[width = 1.15\linewidth]{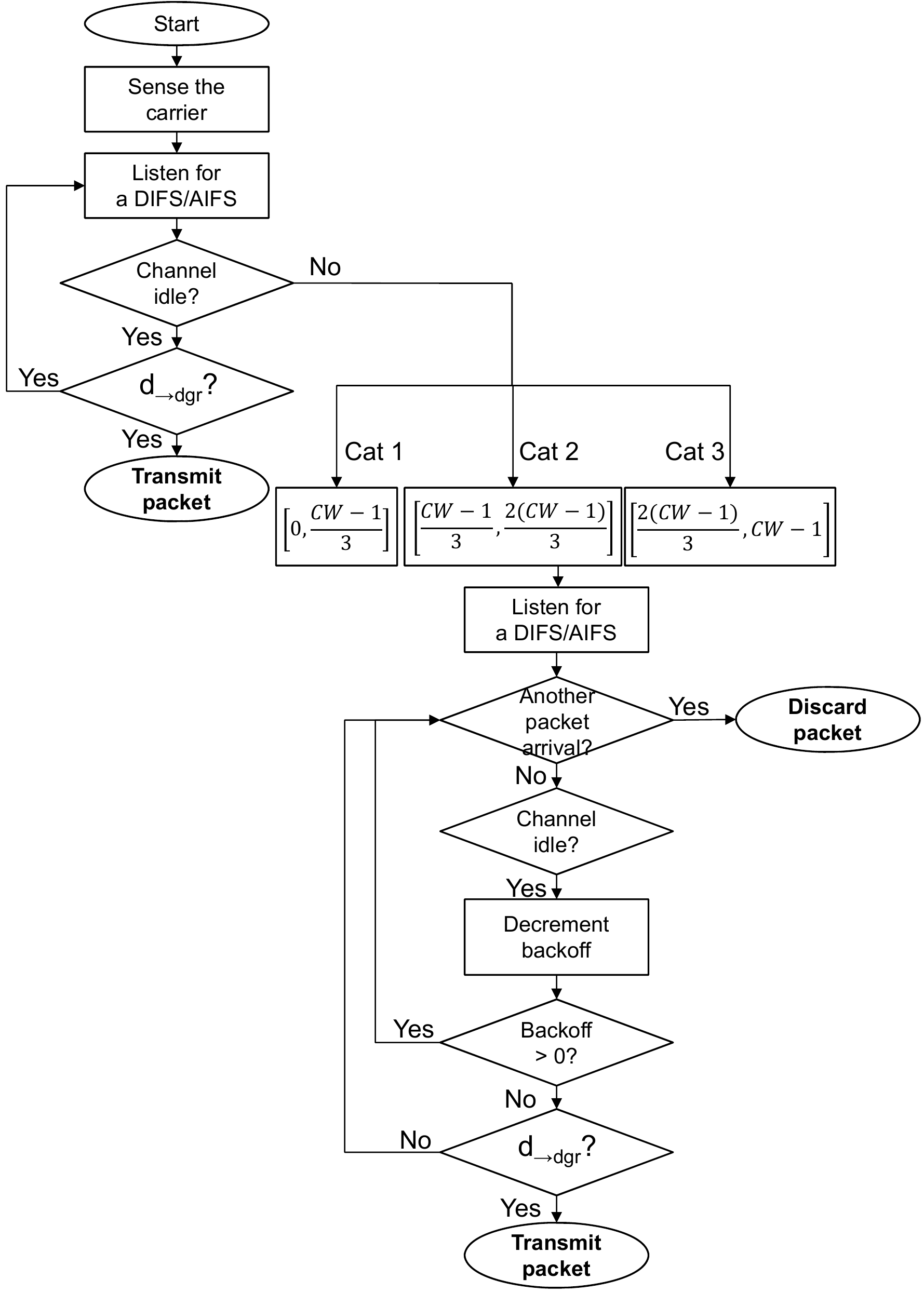}
\caption{Flowchart for the proposed algorithm}
\label{fig_flowchart}
\end{figure}

\subsubsection{Distance Calculation Method}
The foremost key discussion in design of the algorithm is the rationale behind our selection of $d_{\rightarrow\text{dgr}}$ as the metric measuring the risk of a crash. While an accident can be caused by many factors including weather condition, road surface status, mechanical failures, etc, the dominating factor is the inborn reactive time limitation of the drivers \cite{springer13_2}. This makes it reasonable to consider the distance to a danger source as a key factor of an accident \cite{springer13}.

Furthermore, one understands that at the current level of technologies, it is not a difficult task to obtain a vehicle's exact distance from a danger source. Specifically, (i) each BSM contains Global Positioning System (GPS) information; (ii) thus, each vehicle is able to exchange each other's exact position; (iii) as such, each vehicle is able to calculate the distance from each other.

\subsubsection{Backoff Allocation according to $d_{\rightarrow\text{dgr}}$}
Now, based on the aforementioned rationale, we propose \textit{a backoff allocation algorithm according to the distance to a danger}. A flowchart for the proposed mechanism is provided in Figure \ref{fig_flowchart}. Unlike the traditional BEB scheme, the proposed protocol allocates a smaller backoff to the group of vehicles with a smaller $d_{\rightarrow \text{dgr}}$. Specifically, according to the threshold distance, Th$_{i}$, the vehicles in $\mathbb{R}^2$ are grouped in three categories--i.e., CATs 1, 2, and 3. A smaller CAT categorizes a smaller $d_{\rightarrow\text{dgr}}$, which, in turn, means a more urgent need for transmission.

Here is a deeper look into the CATs in relation to a CW. As presented in Section \ref{sec_results}, this paper uses \{300, 500, 700\} m for \{Th$_{1}$, Th$_{2}$, Th$_{3}$\}, representing the thresholds for CATs 1, 2, and, 3, respectively, as have been shown in (\ref{eq_cats}). The proposed protocol divides the entire range of CW into three chunks: for \{Th$_{1}$, Th$_{2}$, Th$_{3}$\}, the backoff counter ranges of [0, (CW-1)/3], [(CW-1)/3, 2(CW-1)/3], and [2(CW-1)/3, CW-1] are allocated. Via this modification, a Tx STA belonging to CAT 1, which is at a higher crash risk due to a shorter $d_{\rightarrow\text{dgr}}$, has to wait for a shorter backoff time. In contrast, a STA with a larger $d_{\rightarrow\text{dgr}}$ is designed to hold a bit longer before a transmission.

\begin{figure*}
\centering
\begin{subfigure}{.495\textwidth}
\centering
\includegraphics[width=\linewidth]{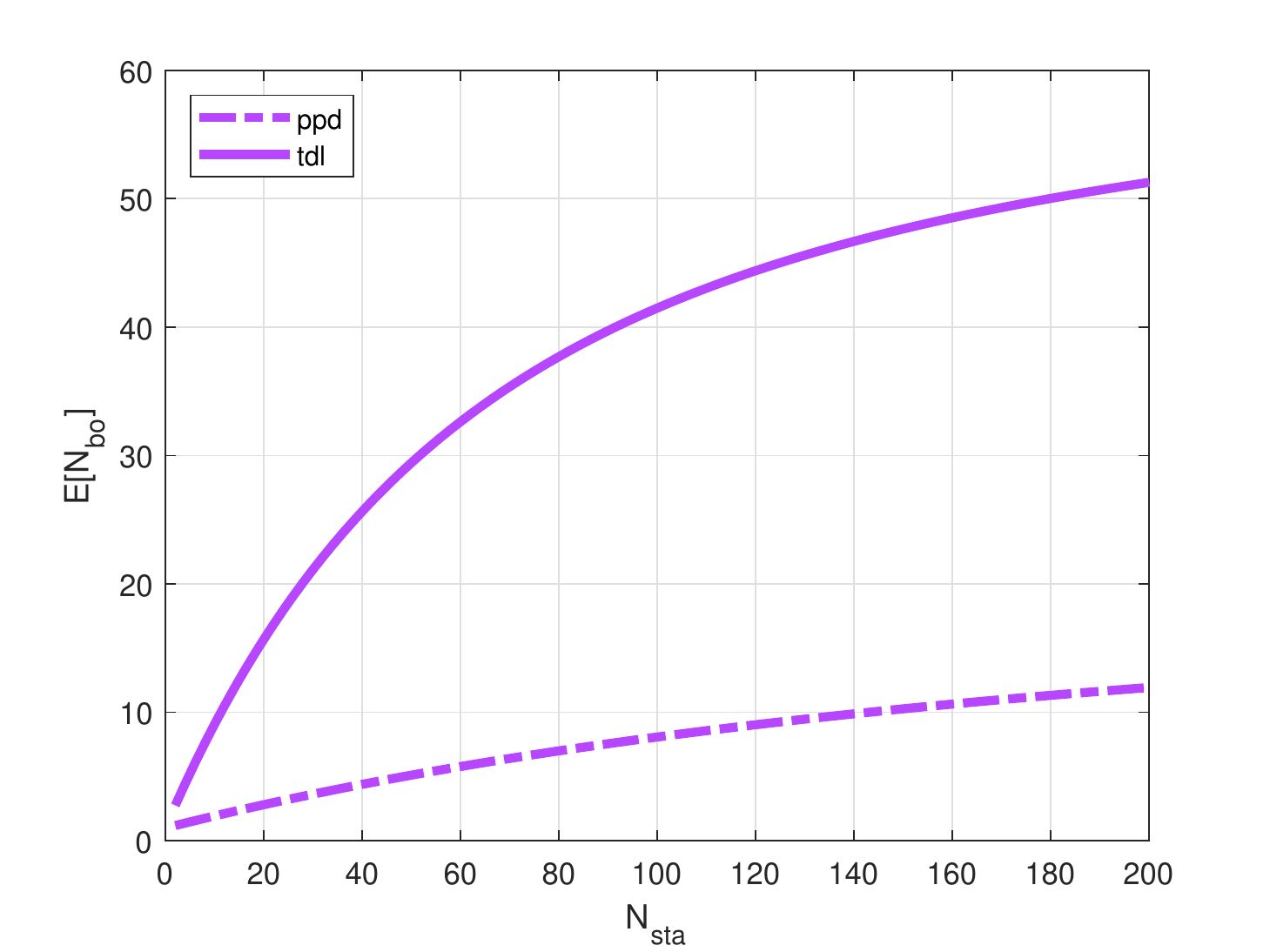}
\caption{Between proposed and traditional schemes (CAT 1 with CW = 127)}
\label{fig_Nbo_schemes}
\end{subfigure}
\begin{subfigure}{.495\textwidth}
\centering
\includegraphics[width=\linewidth]{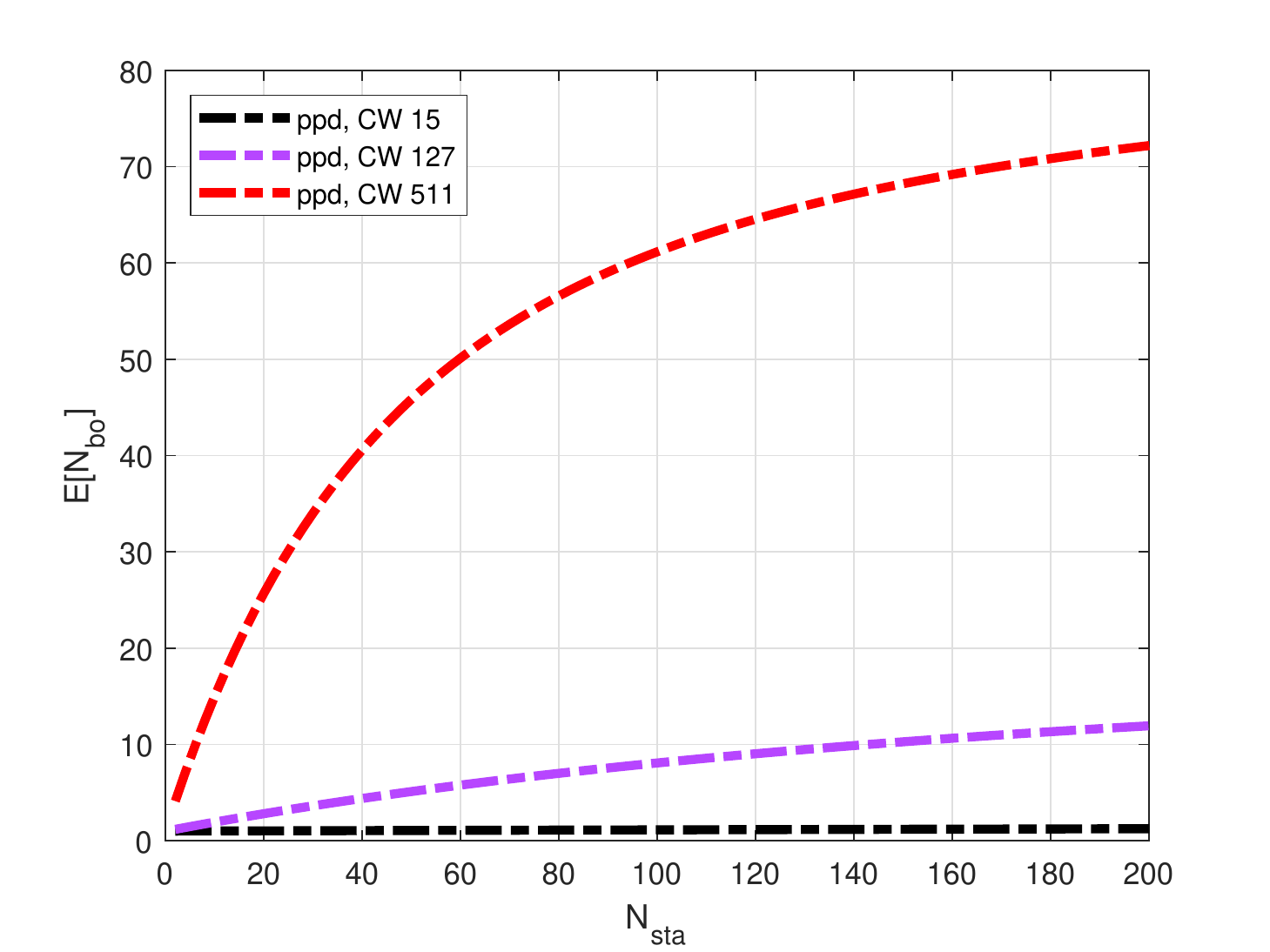}
\caption{According to CW}
\label{fig_Nbo_cw}
\end{subfigure}
\caption{Average number of slots spent during a backoff process vs. number of STAs}
\label{fig_Nbo}
\end{figure*}

%%%%%%%%%%%%%%%%%%%%%%%%%%%%%%%%%%%%%%%%%%%
\section{Performance Analysis}\label{sec_analysis}
This section formulates three metrics to measure the performance of the proposed backoff allocation scheme--namely, average latency, normalized throughput, and inter-reception time $\mathsf{IRT}$.

Moreover, it is worth to notice that all the three quantities are defined with a BSM transmitted at a \textit{tagged vehicle}. We emphasize that such an assumption keeps generality since the type of network being considered in this paper is \textit{distributed}, in which every node has an equal characteristic and hence shows a consistent networking behavior.

\subsection{Average Latency}
We remind that this paper focuses on safety-critical applications, which makes the latency the most significant metric in the performance evaluation of a DSRC network. Further, reflecting the ``broadcast'' nature of a DSRC network, this paper defines an \textit{average latency} among all the STAs across a network.

Let $\mathsf{T}$ denote an instantaneous total latency taken for a node to transmit a packet. Taking into account all the possible results of a packet transmission (i.e., expiration, success, and collision), an average latency can be computed as 
\begin{align}\label{eq_meanT}
\mathbb{E} \big[ \mathsf{T} \big] &= \mathbb{P}\left[\text{Tx}\right]\mathbb{T}\left[\text{Expiration}\right]\nonumber\\
&\hspace{0.4 in}+ \left(1 - \mathbb{P}\left[\text{Tx}\right]\right)\Big\{\mathbb{T}\left[\text{Success}\right] + \mathbb{T}\left[\text{Collision}\right]\Big\}\nonumber\\
&= \left(1 - \tau\right)\mathbb{E}\left[\mathsf{T}_{\text{exp}}\right]\nonumber\\
&\hspace{0.4 in} + \tau\Big\{ \underbrace{\mathbb{E}\left[ \mathsf{T}_{\text{bo}} \right] + \left(1 - \mathsf{P}_{\text{col}}\right)\mathsf{T}_{\text{suc}}}_{=\mathbb{T}\left[\text{Success}\right]} + \mathsf{P}_{\text{col}} \mathsf{T}_{\text{col}}\Big\}\nonumber\\
&\stackrel{(a)}{\approx} \left(1 - \tau\right)\mathbb{E}\left[ \mathsf{T}_{\text{exp}} \right] + \tau \Big\{ \mathbb{E}\left[ \mathsf{T}_{\text{bo}} \right] + \mathsf{T}_{\text{suc}} \Big\}
\end{align}
where $\mathbb{P}\left[\cdot\right]$ and $\mathbb{T}\left[\cdot\right]$ denote the probability and the time length of an event, respectively. Variables in (\ref{eq_meanT}) are defined as follows: $\tau$ denotes the probability that a tagged vehicle is able to transmit in a certain slot within a beaconing period \cite{arxiv19}; $\mathsf{P}_{\text{col}}$ gives the probability of a collision--i.e., a SYNC or a HN \cite{arxiv19}; $\mathsf{T}_{\text{exp}}$, $\mathsf{T}_{\text{col}}$, and $\mathsf{T}_{\text{suc}}$ denote the time lengths taken for an expiration, a collision (i.e., SYNC and/or HN), and a successful delivery, respectfully.

\textit{\textbf{Proof of (a) in (\ref{eq_meanT}):}} {\color{black} Although not presented in this paper, via a separate analysis \cite{arxiv2005}}, the authors approximated as $\mathsf{P}_{\text{col}} \approx 0$ due to a DSRC system being an ``expiration''-constrained system rather than a collision-constrained one. In other words, in a DSRC network, a beaconing period for a BSM is composed of quite a large number of slots (i.e., 1500 slots with a slot time of 66.7 $\mu$sec and a beaconing period of 100 msec), which is beneficial in avoiding a collision while detrimental in being able to transmit before an expiration. (Note that a DSRC BSM expires if it is not transmitted within a beaconing period.)
\hfill$\blacksquare$

\vspace{0.1 in}

Now, each of the terms in (\ref{eq_meanT}) is elaborated in the following proof:

\textit{\textbf{Proof of (\ref{eq_meanT}):}} Starting from the first term, we remind that $\tau$ is the probability of a tagged vehicle being able to go through a backoff process before expiration and thus transmit a packet. For calculation of $\tau$, we modified the Markov chain for a backoff process \cite{bianchi} in order to reflect the impacts of \textit{packet expiration}, which does not occur in classical IEEE 802.11 DCF and hence was not reflected in the existing analysis models for DCF. Due to a long recursiveness in the computation process, it was more efficient to take a numerical approach to obtain $\tau$ instead of a closed-form derivation.

The length of time taken for an expiration, $\mathsf{T}_{\text{exp}}$, can be given by $\mathbb{E}\left[\mathsf{T}_{\text{exp}}\right] = \mathbb{T}\left[\text{beacon}\right]\mathbb{E}\left[\mathsf{N}_{\text{bcn}}\right]$. Notice that the number of consecutive idle beaconing periods, $\mathsf{N}_{\text{bcn}}$, can be characterized as a \textit{geometric random variable} \cite{bianchi}. Based on these formulations, an average time taken for an expiration can be formally written as
\begin{align}\label{eq_Texp}
\mathbb{E}\left[\mathsf{T}_{\text{exp}}\right] &= \mathbb{T}\left[\text{beacon}\right]\mathbb{E}\left[\mathsf{N}_{\text{bcn}}\right]\nonumber\\
&= \mathsf{T}_{\text{ibi}} \cdot \left(1 - \tau\right)\tau^{-1}
\end{align}
where $\mathsf{T}_{\text{ibi}}$ is a constant denoting the inter-broadcast interval.

The second term of (\ref{eq_meanT}) contains the length of time taken for a backoff, which is given by
\begin{align}\label{eq_Tbo}
\mathbb{E}\left[ \mathsf{T}_{\text{bo}} \right] = \mathsf{T}_{\text{slot}} \mathbb{E}\left[ \mathsf{N}_{\text{bo}} \right]
\end{align}
where $\mathsf{T}_{\text{slot}}$ is the length of a slot \cite{bianchi} (i.e., 66.7 $\mu$sec \cite{arxiv19}).

Also, in (\ref{eq_Tbo}), $\mathsf{N}_{\text{bo}}$ denotes the number of slots spent to go through a backoff process. This quantity can be displayed as a function of the number of STAs, denoted by $\mathsf{N}_{\text{sta}}$, as illustrated in Figure \ref{fig_Nbo}. One can observe two main tendencies: (i) from Figure \ref{fig_Nbo_schemes}, the proposed scheme consumes a smaller $\mathsf{N}_{\text{bo}}$ as compared to the traditional CSMA, thanks to higher possibility of shorter backoffs; and (ii) from Figure \ref{fig_Nbo_cw}, a larger CW spends a larger $\mathsf{N}_{\text{bo}}$ due to higher possibility of longer backoffs.

{\color{black}
Lastly, in (\ref{eq_Tbo}), the number of slots that are used by a successful delivery of a packet is formulated as
\begin{align}\label{eq_Tsuc}
\mathsf{T}_{\text{suc}} &= \left(\text{Time for a Data}\right)\nonumber\\
&= \text{Hdr} + \text{Pld} + \text{SIFS} + \mathsf{T}_{\text{prop}}
\end{align}
where Hdr and Pld denote the lengths of a header and a payload, respectively. Also, $\mathsf{T}_{\text{prop}}$ gives the propagation delay, which is assumed to be kept the same to all of the Rx vehicles within the tagged vehicle's communication range.
\hfill$\blacksquare$
}

\subsection{Normalized Throughput}
Based on $\mathbb{E} \big[ \mathsf{T} \big]$ as has been formulated in (\ref{eq_meanT}), we can define the \textit{normalized throughput}, which can be formally written as
\begin{align}\label{eq_throughput}
\mathsf{R} &= \frac{\mathbb{E}\left[ \text{Payload time} \right]}{\mathbb{E}\left[ \text{Time consumed for the payload} \right]}\nonumber\\
&= \frac{\tau \mathsf{T}_{\text{suc}}}{\mathbb{E} \big[ \mathsf{T} \big]}
\end{align}
where $\mathbb{E} \big[ \mathsf{T} \big]$ and $\mathsf{T}_{\text{suc}}$ have already been found in (\ref{eq_meanT}) and (\ref{eq_Tsuc}), respectively. Also, $\tau$ has been mentioned after derivation of (\ref{eq_meanT}) as well.

\subsection{Inter-Reception Time}
Lastly, we define the $\mathsf{IRT}$ as the time taken between two given successful packet reception events. Notice that the unit of a quantity of $\mathsf{IRT}$ is ``the number of beaconing periods.'' As such, one can multiply a beaconing time (e.g., 100 msec in this paper) when wanting to display an $\mathsf{IRT}$ in the unit of time (i.e., seconds).

Now, the probability that $\mathsf{N}_{\text{bcn}}$ failures follow a successful delivery is modeled to follow a geometric distribution, which can be formally written as
\begin{align}\label{eq_irt}
\mathsf{IRT} = \left(1 - \tau\right)^{\mathsf{N}_{\text{bcn}} - 1} \tau.
\end{align}

\textit{\textbf{Proof of (\ref{eq_irt}):}} For occurrence of an ``$\mathsf{IRT}$,'' we suppose to start from a successful reception, and then measure how many beaconing periods are expended until the next successful reception. That is, the first beaconing period is set to have the probability of $\mathsf{P}_{\text{suc}}$, and thereafter the possibility is left open between $\mathsf{P}_{\text{suc}}$ and $1 - \mathsf{P}_{\text{suc}}$ depending on occurrence of a successful delivery or a failure, respectfully. This can be formulated as
\begin{align}
&\mathsf{IRT} \sim \text{Geo}\left(\mathsf{P}_{\text{suc}}\right)\nonumber\\
&\hspace{0.4 in}\Rightarrow \mathbb{P}\left[\mathsf{IRT} = \mathsf{N}_{\text{bcn}}\right] = \left(1 - \mathsf{P}_{\text{suc}}\right)^{\mathsf{N}_{\text{bcn}}-1} \mathsf{P}_{\text{suc}}\nonumber\\
&\hspace{1.35 in}\stackrel{(a)}{\approx} \left(1 - \tau\right)^{\mathsf{N}_{\text{bcn}}-1} \tau
\end{align}
where (a) follows from the fact that $\mathsf{P}_{\text{suc}}$ gives the probability of a successful reception, which is given by
\begin{align}
\mathsf{P}_{\text{suc}} = \tau \left(1 - \mathsf{P}_{\text{col}}\right) \approx \tau
\end{align}
since $\mathsf{P}_{\text{col}} \approx 0$ as has already been mentioned in Proof of (a) in (\ref{eq_meanT}).
\hfill$\blacksquare$

\begin{table}[t]
\caption{Values for key parameters}
\centering
\begin{tabular}{ |l|l|}
\hline
\textbf{\cellcolor{gray!30}Parameter (Symbol)} & \cellcolor{gray!30}\textbf{Value}\\
\hline\hline
Inter-broadcast interval ($\mathsf{I}_{\text{ibi}}$) & 100 msec\\
DIFS& 128 $\mu$s\\
SIFS& 28 $\mu$s\\
Payload length (Pld) & 40 bytes \cite{j2735}\\
Propagation delay $(\mathsf{T}_{\text{prop}})$ & 1 $\mu$s\\
Slot time $\left(\mathsf{T}_{\text{slot}}\right)$ & 50 $\mu$s\\
RTS & 300 $\mu$s\\ 
ACK & 300 $\mu$s\\ 
CTS & 350 $\mu$s\\
Space size $\left(\left|\mathbb{R}^2\right|\right)$ & 2 km by 2 km\\
Vehicle density $\left(\lambda\right)$ & $2 \times 10^{-5}$ m$^{-2}$\\
Cat threshold distance (\{Th$_{1}$, Th$_{2}$, Th$_{3}$\}) & \{300, 500, 700\} m\\
\hline
\end{tabular}
\label{table_parameters}
\end{table}

\begin{figure*}
\centering
\begin{subfigure}{\textwidth}
\centering
\includegraphics[width=\linewidth]{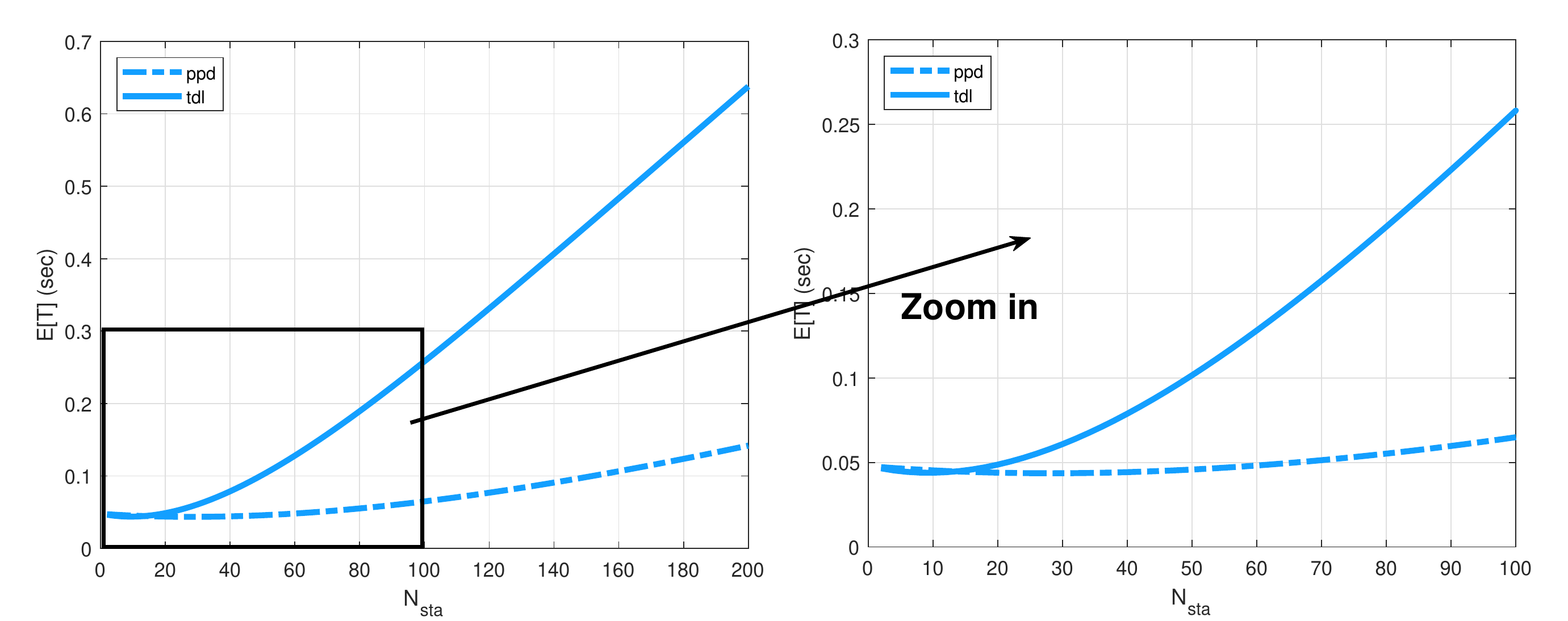}
\caption{{\color{black}Between proposed and traditional schemes (CAT 1 with CW = 127)}}
\label{fig_T_schemes}
\vspace{0.2 in}
\end{subfigure}
\begin{subfigure}{\textwidth}
\centering
\includegraphics[width=\linewidth]{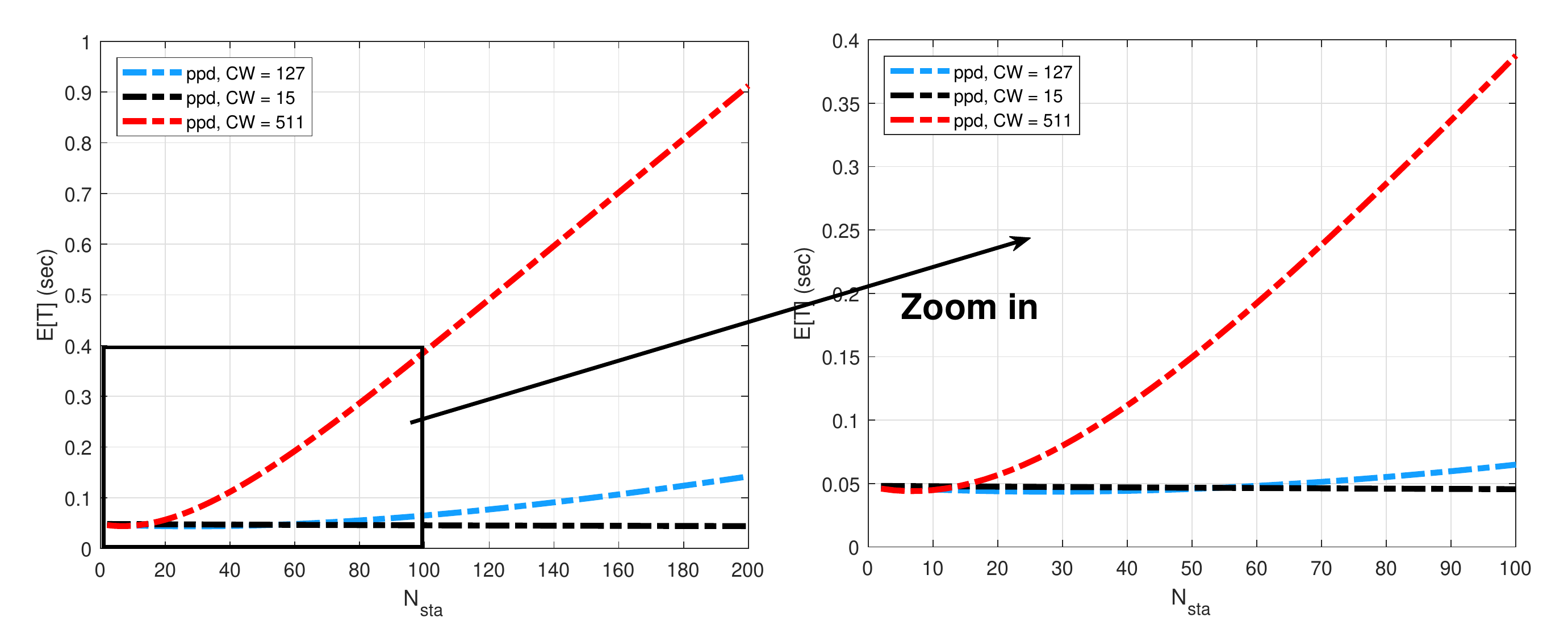}
\caption{{\color{black}According to CW}}
\label{fig_T_cw}
\end{subfigure}
\caption{{\color{black}Average delay vs. number of STAs}}
\label{fig_T}
\vspace{0.2 in}
\end{figure*}

%%%%%%%%%%%%%%%%%%%%%%%%%%%%%%%%%%%%%%%%%%%
\section{Numerical Results}\label{sec_results}
In this section, we evaluate the performance of the proposed backoff algorithm compared to the traditional CSMA (i.e., BEB) \cite{11pstd}. As summarized in Table \ref{table_parameters}, each Tx STA is assumed to have a fixed payload length of 40 bytes \cite{j2735}. Also, for our numerical analysis, we set the spatial setting being consistent with what has been shown in Figure \ref{fig_scatter}: (i) 80 vehicles were placed following a uniform distribution with respect to both $X$ and $Y$ in $\mathbb{R}^2$ (i.e., $\lambda = 2 \times 10^{-5}$ m$^{-2}$); (ii) a danger source at the origin; and (iii) the size of $\mathbb{R}^2$ is 2 km by 2 km.

\subsection{Average Packet Delay}
Figure \ref{fig_T} demonstrates the average delay, $\mathbb{E}\left[\mathsf{T}\right]$, versus the number of STAs, $\mathsf{N}_{\text{sta}}$, according to the Cat and CW. We remind from (\ref{eq_meanT}) that $\mathbb{E}\left[\mathsf{T}\right]$ is composed of two parts: i.e., (i) $\mathsf{T}_{\text{exp}}$ and (ii) $\mathbb{E}\left[\mathsf{T}_{\text{bo}}\right] + \mathsf{T}_{\text{suc}}$. As $\mathsf{N}_{\text{sta}}$ grows, $\left(1 - \tau \right)$, $\mathsf{T}_{\text{exp}}$, and $\mathsf{T}_{\text{bo}}$ increase while $\mathsf{T}_{\text{suc}}$ remains as a constant.

This explains why the quantity of $\mathbb{E}\left[\mathsf{T}\right]$ forms an \textit{inflection point} as shown in every curve in Figures \ref{fig_T_schemes} and \ref{fig_T_cw}. To highlight formation of an inflection point, each of Figures \ref{fig_T_schemes} and \ref{fig_T_cw} is zoomed as shown on the right-hand side.

\begin{figure*}
\centering
\begin{subfigure}{.495\textwidth}
\centering
\includegraphics[width=\linewidth]{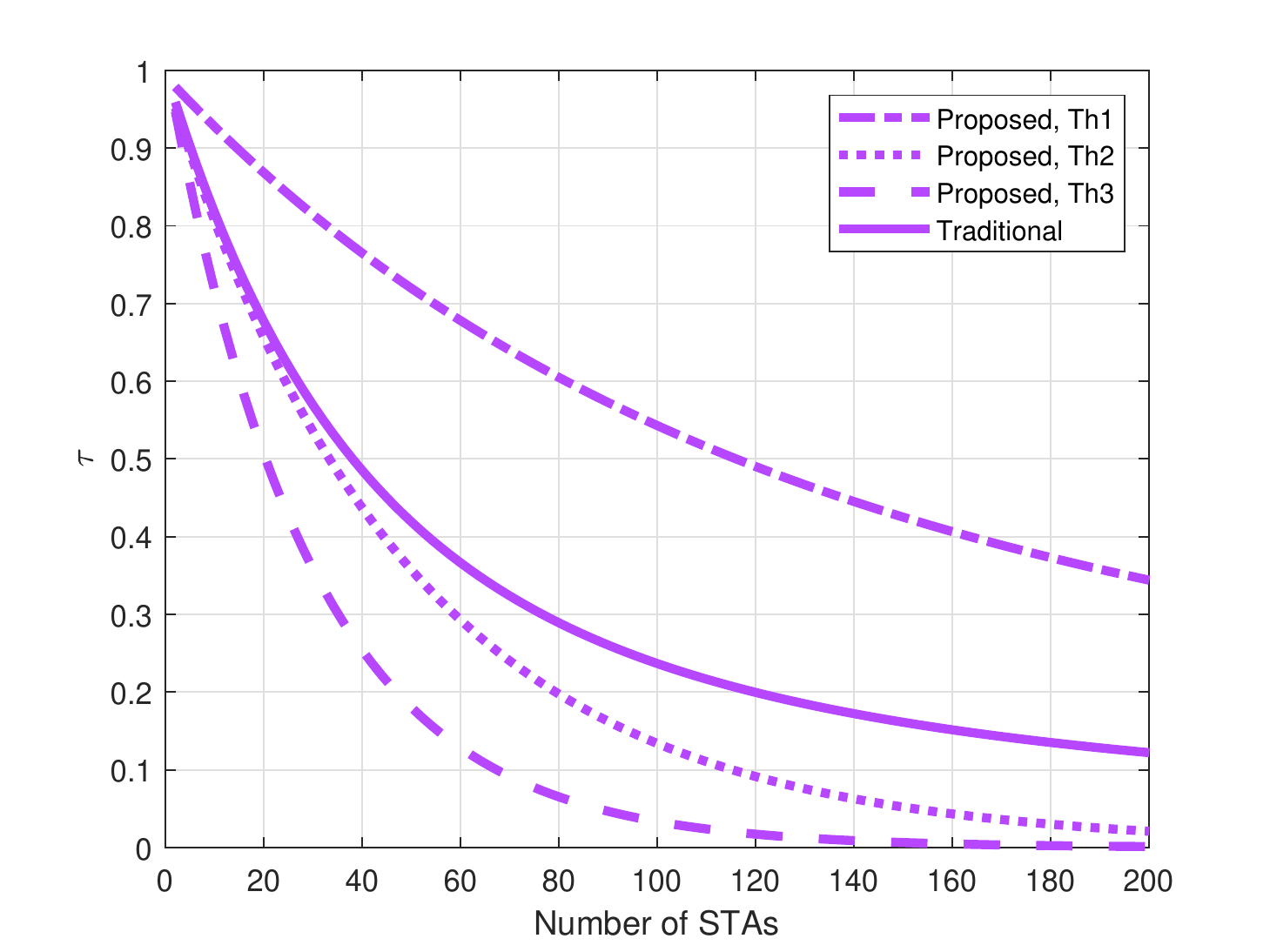}
\caption{Between proposed and traditional schemes (CAT 1 with CW = 127)}
\label{fig_tau_schemes}
\end{subfigure}
\begin{subfigure}{.495\textwidth}
\centering
\includegraphics[width=\linewidth]{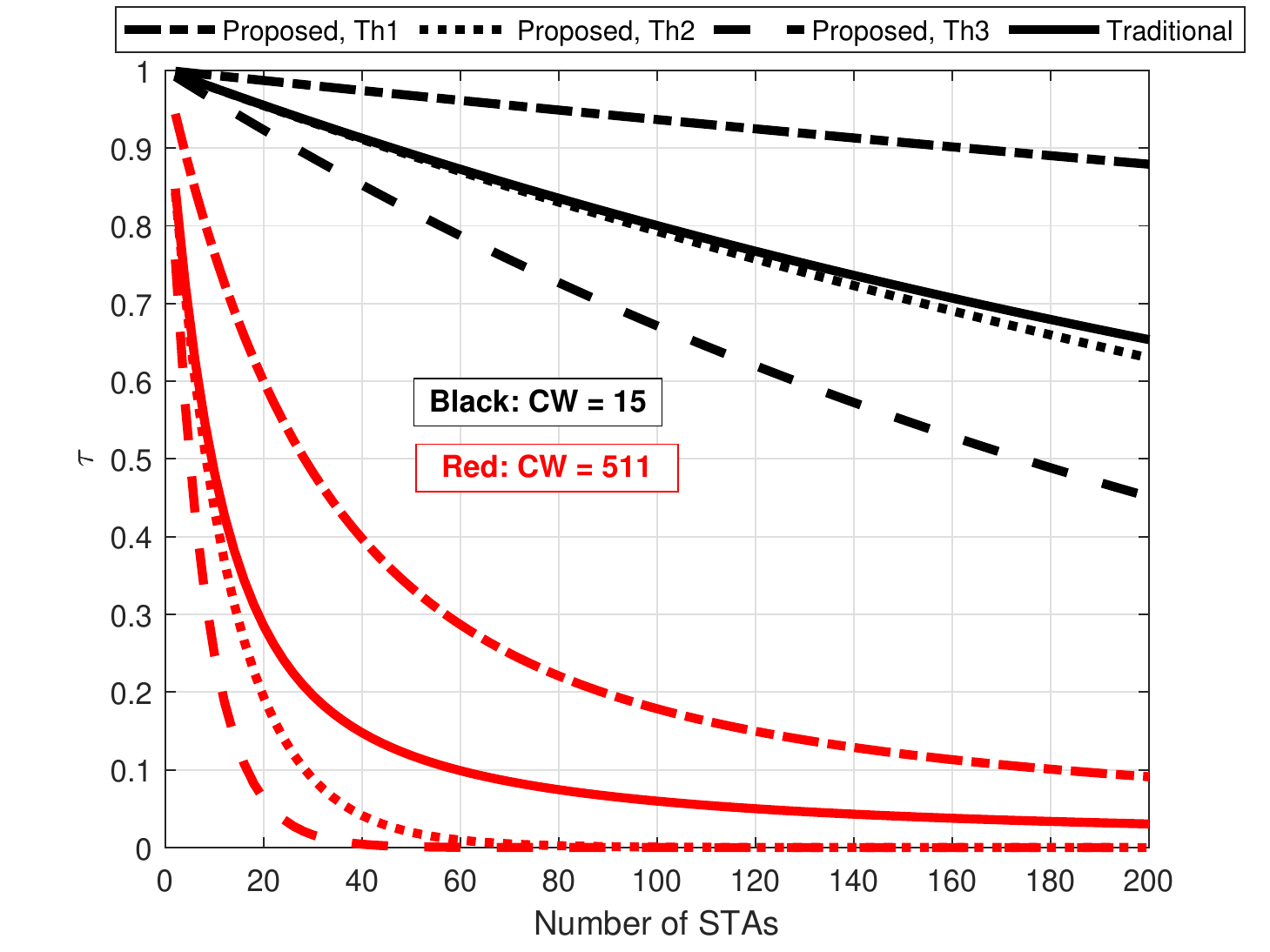}
\caption{According to CW}
\label{fig_tau_cw}
\end{subfigure}
\caption{Probability of packet transmission vs. number of STAs}
\label{fig_tau}
\vspace{0.2 in}
\end{figure*}

Let us start from $\mathsf{N}_{\text{sta}} = 0$ and increase to investigate the tendency of $\mathbb{E}\left[\mathsf{T}\right]$. Specifically, the tendency will decrease at the beginning, bounce back forming an inflection point, and proceed to increasing further thereafter. The quantitative rationale is given as below:
\begin{itemize}
\item Although not explicitly displayed in the figure, we discovered that initially (i.e., with $\mathsf{N}_{\text{sta}} \approx 0$), $\mathsf{T}_{\text{exp}} << \mathsf{T}_{\text{bo}}+\mathsf{T}_{\text{suc}}$ and the weight for $\mathsf{T}_{\text{bo}}+\mathsf{T}_{\text{suc}}$ is greater than that for $\mathsf{T}_{\text{exp}}$ since $\tau$ is large with a small $\mathsf{N}_{\text{sta}}$, as shown in Figure \ref{fig_tau}.
\item Soon after, with $\mathsf{N}_{\text{sta}}$ being still relatively small, we found that while still $\mathsf{T}_{\text{exp}} < \mathsf{T}_{\text{bo}}+\mathsf{T}_{\text{suc}}$, $\mathsf{T}_{\text{exp}}$ grows far faster than $\mathsf{T}_{\text{bo}}+\mathsf{T}_{\text{suc}}$. Now, the weight for $\mathsf{T}_{\text{exp}}$ is greater than tha for $\mathsf{T}_{\text{bo}}+\mathsf{T}_{\text{suc}}$ since $\tau$ gets smaller due to $\mathsf{N}_{\text{sta}}$ being greater.
\item Eventually, however, with $\mathsf{N}_{\text{sta}}$ being large enough, both the quantity and weight get bigger in the first term: i.e., $\mathsf{T}_{\text{exp}} > \mathsf{T}_{\text{bo}}+\mathsf{T}_{\text{suc}}$ and $\left(1 - \tau\right) > \tau$ as $\tau$ gets far smaller due to $\mathsf{N}_{\text{sta}}$ getting very large.
\end{itemize}

This tendency also explains why the quantity of $\mathbb{E}\left[\mathsf{T}\right]$ ranges very large (i.e., to almost 1 sec) even with the proposed scheme (with CW = 511) as shown in Figure \ref{fig_T_cw}. In addition to the fact that $\mathsf{T}_{\text{exp}}$ grows faster than $\mathbb{E}\left[\mathsf{T}_{\text{bo}}\right] + \mathsf{T}_{\text{suc}}$ as $\mathsf{N}_{\text{sta}}$ gets larger, the weight, i.e., $\left(1 - \tau \right)$, also gets greater. This relationship leads the resulting average delay, $\mathbb{E}\left[\mathsf{T}\right]$, to such a large number.

\subsection{Normalized Throughput}
\begin{figure*}
\centering
\begin{subfigure}{.495\textwidth}
\centering
\includegraphics[width=\linewidth]{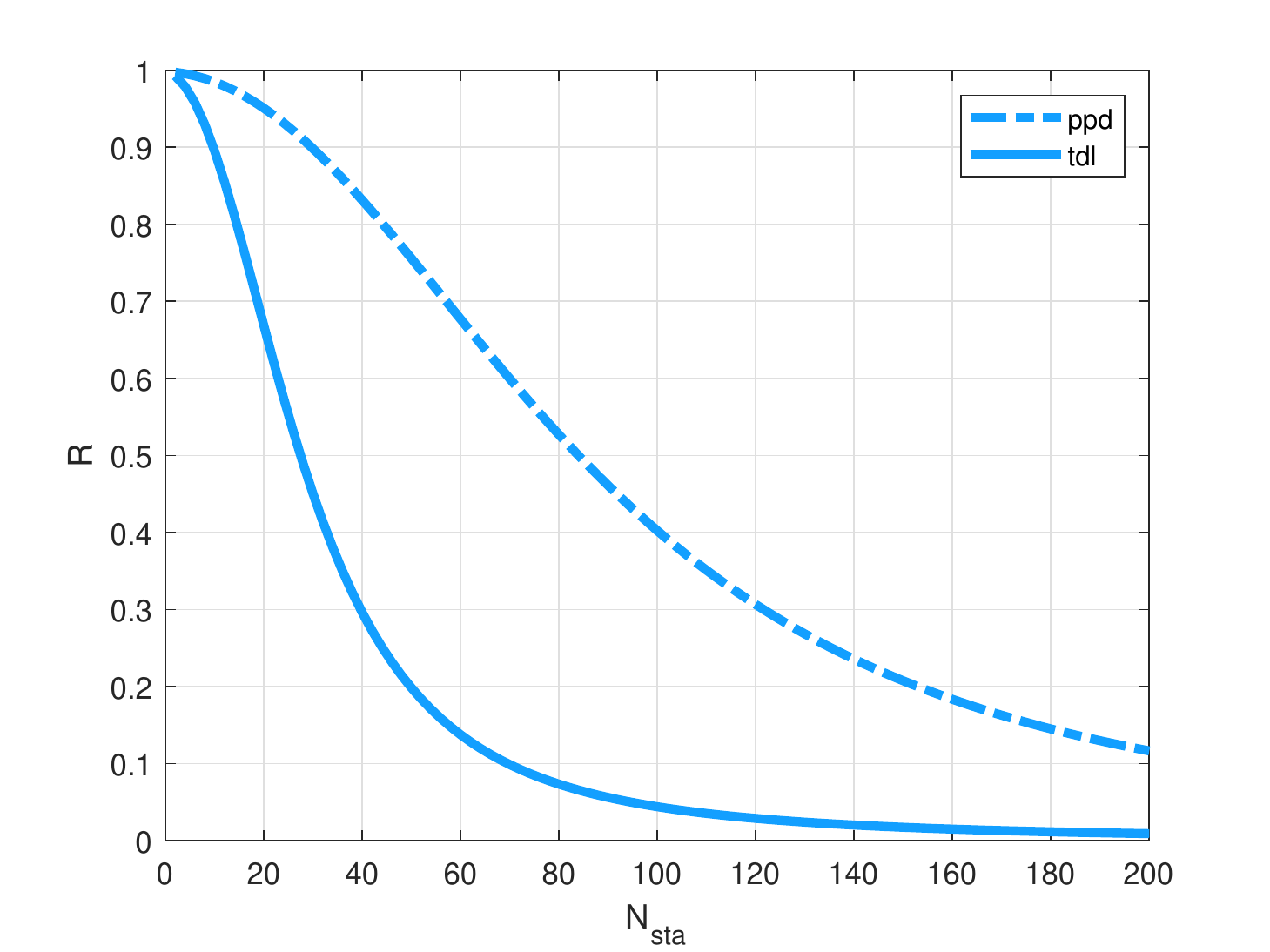}
\caption{{\color{black}Between proposed and traditional schemes (CAT 1 with CW = 127)}}
\label{fig_throughput_schemes}
\end{subfigure}
\begin{subfigure}{.495\textwidth}
\centering
\includegraphics[width=\linewidth]{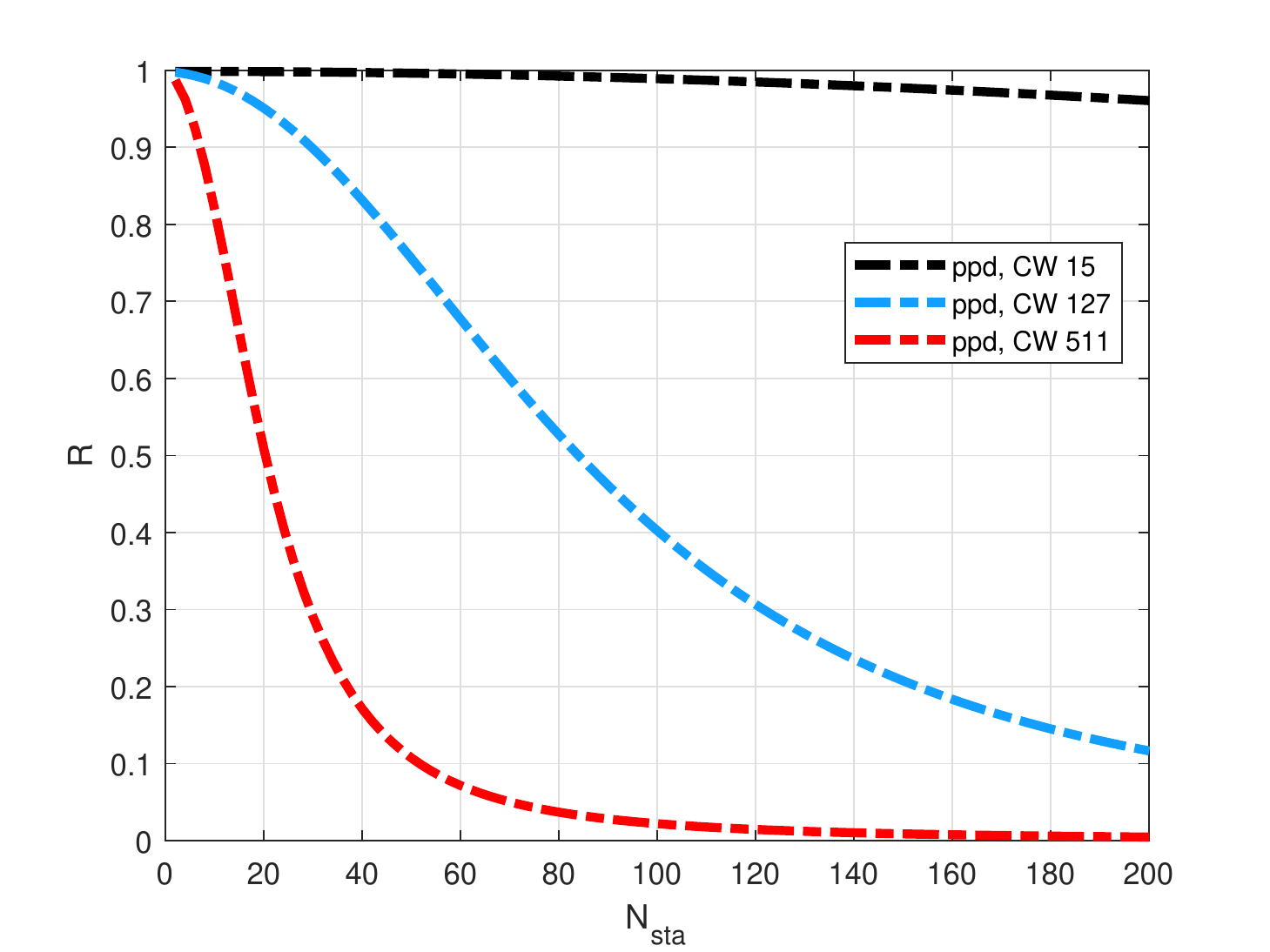}
\caption{{\color{black}According to CW}}
\label{fig_throughput_cw}
\end{subfigure}
\caption{{\color{black}Normalized throughput vs. number of STAs}}
\label{fig_throughput}
\end{figure*}

As has been shown in (\ref{eq_throughput}), for analysis of the normalized throughput, $\mathsf{R}$, the key variables to characterize are $\tau$ and $\mathbb{E}\left[\mathsf{T}\right]$, while $\mathsf{T}_{\text{suc}}$ has been given as a constant as formulated in (\ref{eq_Tsuc}). Hence, here we quantify $\tau$ and $\mathbb{E}\left[\mathsf{T}\right]$, which are displayed in Figures \ref{fig_tau} and \ref{fig_throughput}.

Figure \ref{fig_tau} presents $\tau$, the probability that a STA transmits in an arbitrary slot, versus $\mathsf{N}_{\text{sta}}$, the number of STAs competing for the medium. The figure also demonstrates comparison according to the threshold distance and CW.

Figure \ref{fig_tau_schemes} compares $\tau$ between the proposed and traditional CSMA schemes for the most dangerous vehicles (i.e., CAT 1 from Figure \ref{fig_scatter}) with CW = 127. It is straightforward that a larger $\mathsf{N}_{\text{sta}}$ causes a lower $\tau$. Also, the level of $\tau$ is significantly improved in the proposed scheme in comparison to the traditional CSMA. It highlights the principle of the proposed scheme: higher chances of transmissions in a backoff process are granted the vehicles that are closer to the danger source where \textit{semantic message prioritization} is accomplished.

Figure \ref{fig_tau_cw} compares $\tau$ according to CW. Commonly with the proposed and traditional schemes, a larger CW results in a lower $\tau$ due to a longer backoff process. From this, one can infer that a DSRC network is a ``expiration-constrained'' network rather than a ``collision-constrained'' one; if it was a collision-constrained, a larger CW would have yielded a higher performance.

Now, Figure \ref{fig_throughput} gives $\mathsf{R}$ versus $\mathsf{N}_{\text{sta}}$ according to the threshold distance. We remind from (\ref{eq_throughput}) that $\mathsf{R}$ is directly proportional to $\tau$, which yields that Figure \ref{fig_throughput} shows a similar overall tendency to what Figure \ref{fig_tau} did.

\begin{figure*}
\centering
\begin{subfigure}{.495\textwidth}
\centering
\includegraphics[width=\linewidth]{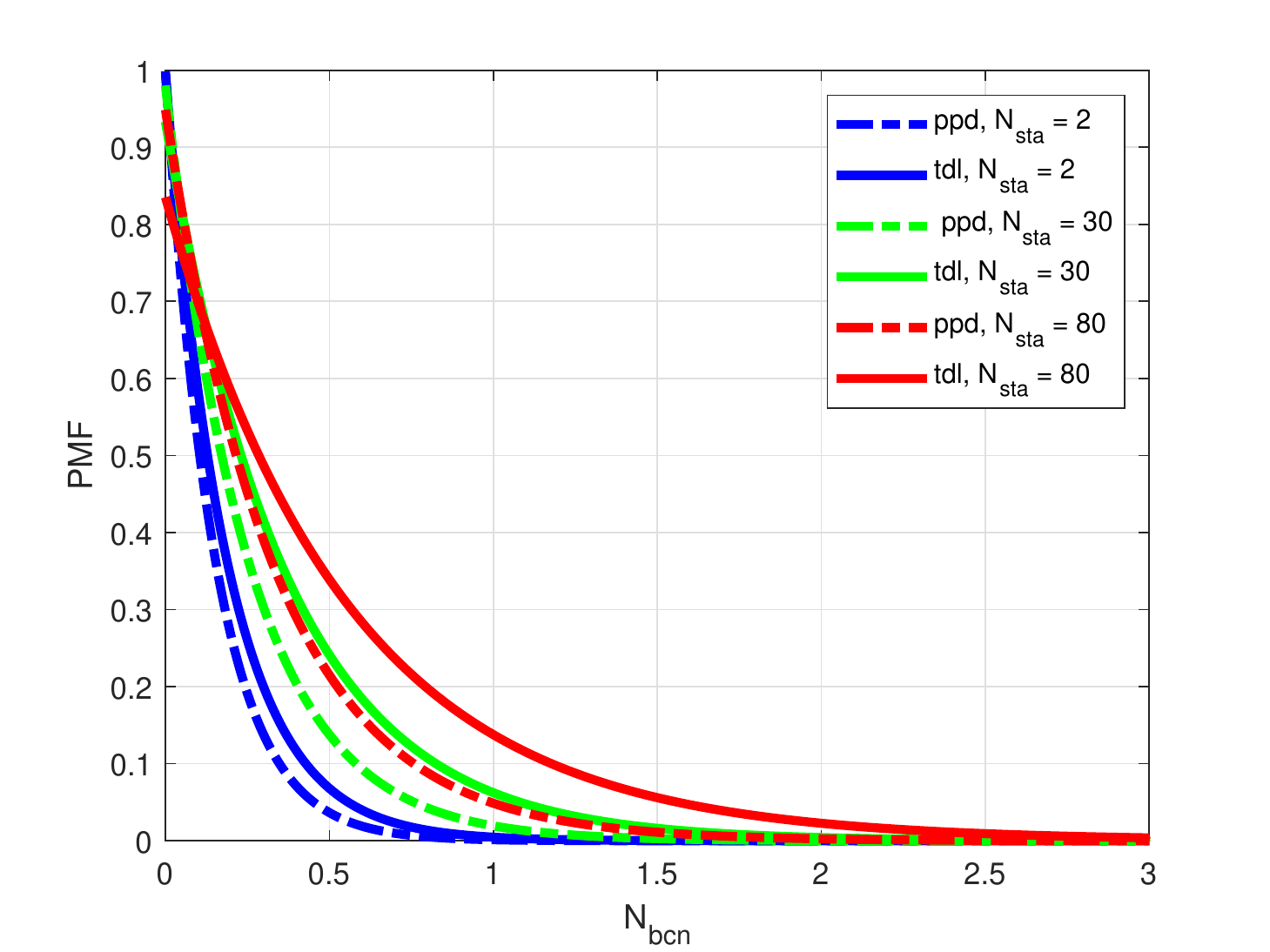}
\caption{PMF}
\label{fig_IRT_PMF}
\end{subfigure}
\begin{subfigure}{.495\textwidth}
\centering
\includegraphics[width=\linewidth]{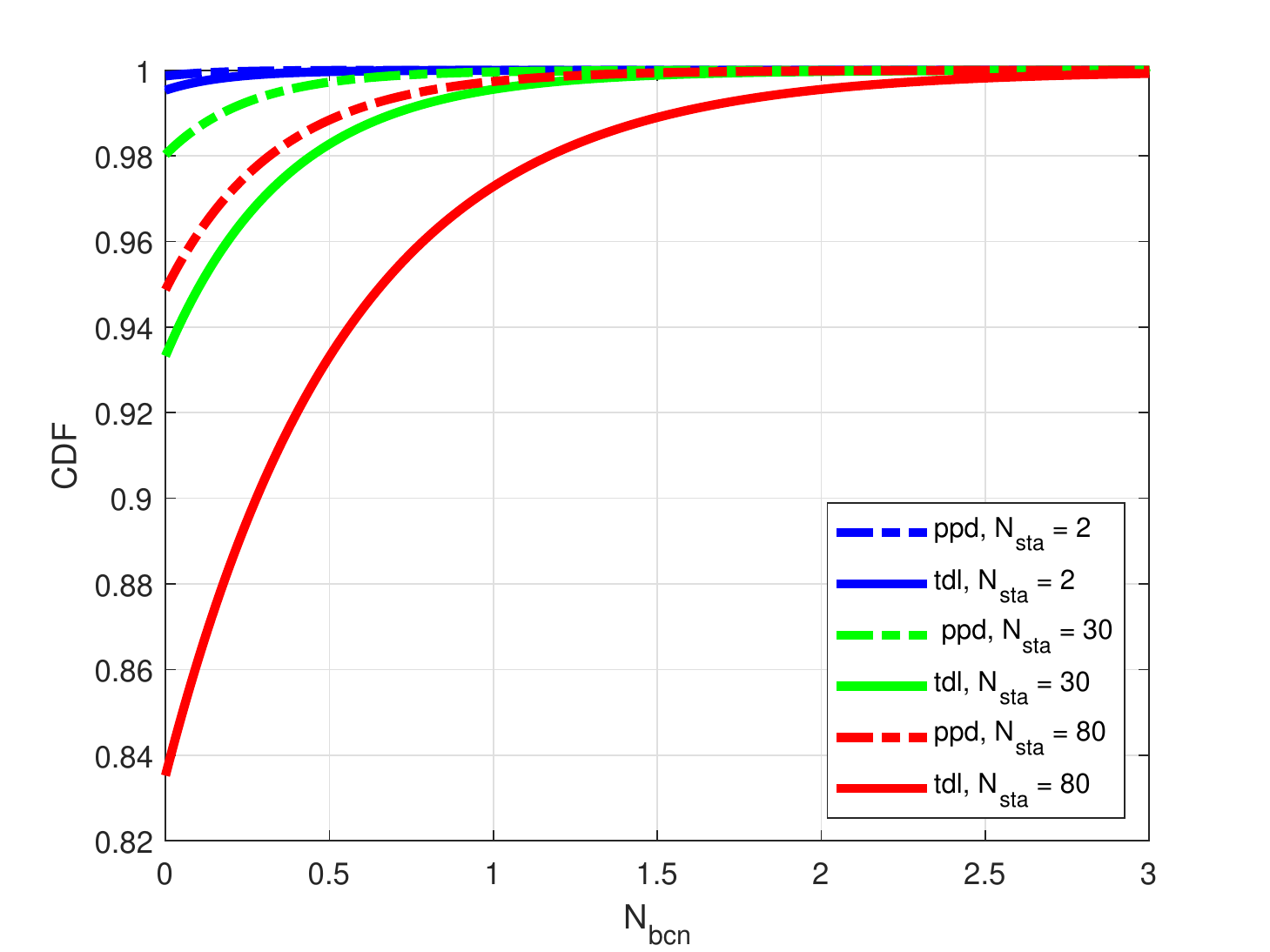}
\caption{CDF}
\label{fig_IRT_CDF}
\end{subfigure}
\caption{Distribution of $\mathsf{IRT}$ (CAT 1 with CW = 15)}
\label{fig_IRT}
\end{figure*}

\subsection{Inter-Reception Time}
Figures \ref{fig_IRT_PMF} and \ref{fig_IRT_CDF} present the probability mass function (PMF) and cumulative distribution function (CDF) of random variable $\mathsf{N}_{\text{bcn}}$ (with CW = 15), respectfully. Each of the figures demonstrate comparisons according to (i) whether the proposed or traditional scheme and (ii) $\mathsf{N}_{\text{sta}}$ and (ii) $\mathsf{N}_{\text{sta}}$.

The figures describe that commonly in observation of PMF and CDF, zero to one beaconing period is consumed during a packet transmission failure. Notice that $\mathsf{N}_{\text{bcn}} = 0$ means a successful packet delivery in the first attempt, without any failure before the successful delivery. In this paper's system model, each failed transmission takes an additional 100 msec since an entire beaconing period is wasted without transmitting a BSM, which translates the results to \textit{consumption of 0 to 100 msec of} $\mathsf{IRT}$.

Moreover, both of Figures \ref{fig_IRT_PMF} and \ref{fig_IRT_CDF} suggest that the overall tendencies in regards to $\mathsf{N}_{\text{sta}}$ and the type of scheme are consistent with the other two metrics: (i) a $\mathsf{N}_{\text{sta}}$ leads to a higher probability of experiencing a shorter $\mathsf{IRT}$; (ii) the proposed scheme outperforms the traditional one, by yielding a shorter $\mathsf{IRT}$, regardless of $\mathsf{N}_{\text{sta}}$. 

The figures also suggest that the superiority of the proposed protocol gets greater with more vehicles competing for the medium. This serves as another concrete evidence that a DSRC network is more expiration-constrained rather than collision-constrained.

%%%%%%%%%%%%%%%%%%%%%%%%%%%%%%%%%%%%%%%%%%%%%%%%%%%%%
\section{Conclusions}
This paper proposed a protocol prioritizing the transmission of a BSM for a vehicle with a higher level of accident risk. Our results showed that this protocol effectively improved the performance of vehicles with higher risk, measured in terms of key metrics--i.e., average delay, throughput, and inter-reception time ($\mathsf{IRT}$). This paper also provided a generalized (i) analytical framework and (ii) spatial system model for evaluating the performance of the proposed scheme according to key factors such as the number of competing STAs and CW.

Thanks to the generality, this work can be extended in multiple directions. For instance, based on the general model of node distribution (as opposed to previous work limiting the models to ``road'' environments), this paper's findings can be applied to other types of transportation network such as unmanned aerial vehicles (UAVs) for building a stochastic geometry-based framework analyzing delay and throughput performances.

{\color{black}
It will be a meaning attempt to extend this work if one (i) considers multiple factors potentially causing an accident and (ii) finds an explicit relationship among them to quantify the accident risk. For instance, it will be easy to identify a number of risk factors; but the hard part will be to characterize the exact impact on the accident risk as a result of the factors in concert.
}

%%%%%%%%%%%%%%%%%%%%%%%%%%%%%%%%%%%%%%%%%%%%%%%%%%%%%%%%

\end{document}